\documentclass[useAMS,usenatbib,referee]{mn2e}
\usepackage{bm}
\input epsf.tex
\usepackage{graphicx}
\def\b{\bm}
\def\oline{\overline}
\title[Formulation of the solar surface dynamo]
       {Formulation of the solar surface dynamo}

\author[V. Krishan]{V. Krishan$^{1,2}$\thanks{E-mail:
vinod@iiap.res.in} \\
$^1$Indian Institute of Astrophysics, Bangalore 560034, India\\ 
$^2$Raman Research Institute, Bangalore 560080, India}

\begin{document}

\date{Accepted--\hskip 1 cm   Received in original form --}

\pagerange{\pageref{firstpage}--\pageref{lastpage}} \pubyear{2008}

\maketitle

\label{firstpage}

\begin{abstract}
The solar surface dynamo has become an active area of research in an attempt to understand the origin of a variety of magnetic field structures on the sun. The major modification that needs to be incorporated in the standard dynamo process is the inclusion of the partial ionization of the gas in the layers underlying and overlaying the photosphere along with the effects associated with the multifluid nature of the system. This not only changes the inertia carrying species but also substantially modifies the temporal and spatial evolution of the magnetic induction. The energy equation also carries the import of these non-ideal effects. The effects such as the Hall effect and the ambipolar diffusion take the dynamo study beyond the realm of the ideal magnetohydrodynamics. In this paper, a first principle formulation of the solar surface dynamo problem has been attempted.  
\end{abstract}


\section{Introduction}
 V\"{o}gler and Sch\"{u}ssler ( 2007) with their paper on solar surface dynamo have opened up a new area of the investigation of the solar magnetic field structures. The solar surface fields on relatively short spatial and time scales with varying geometries are now observed and inferred routinely. This necessitates the inclusion of the realistic solar atmospheric models in the dynamo mechanism.  V\"{o}gler and Sch\"{u}ssler ( 2007) have studied the surface dynamo including the effects of stratification, compressibility, partial ionization, radiative transfer as well as an open lower boundary. While this is a major effort, the magnetic induction equation they have used is still bereft of the effects ensuing from the multifluid interactions in a partially ionized plasma specifically the Hall and the ambipolar diffusion effects. Recently Krishan and Gangadhara (2008) initiated a study of the meanfield dynamo in a partially ionized plasma including the Hall and the ambipolar effects in the induction equation for closed boundary conditions while neglecting stratification, compressibility and radiative transfer effects. In this paper an attempt is made to formulate the solar surface dynamo including the Hall effect and the ambipolar diffusion while retaining all the effects included in  V\"{o}gler and Sch\"{u}ssler ( 2007). A single fluid description of the partially ionized plasma in the limit of weak ionization is presented in section 2. The mean-field dynamo equation in the turbulent partially ionized plasma is derived in section 3 and the expressions for the $\alpha$ and the $\beta$ parameters of a meanfield dynamo are extracted. A crude estimate of the time scale of evolution of the magnetic field due to turbulent ambipolar diffusion arising due to the fluctuating ion-neutral collision frequency is given.

\section{Three-component magnetofluid}

We begin with the three component partially ionized plasma consisting
of electrons (e), ions (i) of mass density $\rho_i$ and
neutral particles (n) of mass density $\rho_n$. A weakly ionized plasma is
defined by the condition (Alfv\'{e}n and F\"althammer 1962) that the
electron-neutral collision frequency $\nu_{\rm en}\sim 10^{-15}n_{\rm
  n}\sqrt{8K_{\rm B}T/(\pi m_{\rm en})}$ is much larger than the
electron-ion collision frequency $\nu_{\rm ei}\sim 6\times
10^{-24}n_{\rm i}\Lambda Z^2(K_{\rm B}T)^{-3/2}$. This translates into
the ionization fraction $n_{\rm e}/n_{\rm n} < 5\times 10^{-11}T^2$
(Alfv\'{e}n and F\"althammer 1962) where $n$'s are the particle
densities and $T$ is the temperature in Kelvin. A major part of the
solar photosphere (Leake \& Arber 2006; Krishan \& Varghese 2007) qualifies to be a weakly ionized plasma.
The equation
of motion of the electrons can be written as:
\begin{eqnarray} 
&&m_{\rm e} n_{\rm e}\left[\frac{\partial {\b V}_{\rm e}}{\partial t}+ 
({\b V}_{\rm e}\cdot\nabla){\b V}_{\rm e}\right]=
 -\nabla p_{\rm e}- \nonumber \\
&&\quad\quad e n_{\rm e}\left[{\b E}+
 \frac{\b V_{\rm e}\times \b B}{c}\right]-
 m_{\rm e}n_{\rm e}\nu_{\rm en}(\b V_{\rm e}-\b V_{\rm n})~.
\end{eqnarray} 
where the electron-ion collisions have been neglected since the
ionized component is of low density.  On neglecting the electron
inertial force, the electric field $\b E$ is found to be:
\begin{eqnarray} 
{\b E}=-\frac{\b V_{\rm e}\times \b B}{c}-\frac{\nabla
    p_{\rm e}}{en_{\rm e}}-\frac{m_{\rm e}}{e}\nu_{\rm en}(\b V_{\rm e}-\b V_{\rm n})~.
\end{eqnarray}
This gives us Ohm's law. For $\delta=(\rho_{\rm i}/\rho_{\rm n}) \ll 1$ along with the dominance of the ion-neutral collisional force over the inertial force, the ion
dynamics can be ignored. The ion force balance then becomes:
\begin{eqnarray}
0=-\nabla p_{\rm i}+e n_{\rm i}\left[{\b E}+\frac
{\b V_{\rm i}\times \b B}{c}\right]-\nu_{\rm in}\rho_{\rm i}(\b V_{\rm i}-\b V_{\rm n})~,
\end{eqnarray}
where $\nu_{\rm in} $ is the ion-neutral collision frequency, and the
ion-electron collisions have been neglected for the low density
ionized component.  Substituting for $\b E$ from Eq.~(2) we find the
relative velocity between the ions and the neutrals:
\begin{equation}
\b V_{\rm n}-\b V_{\rm i}= \frac{\nabla (p_{\rm i}+p_{\rm e})}{\nu_{\rm in} 
\rho_{\rm i}}-\frac{\b J\times \b B}{ c\nu_{\rm in}\rho_{\rm i}}-\frac{m_{\rm e}\nu_{\rm en}}{e}\b J, 
\end{equation}
where
\begin{equation}
\b J = en_{\rm e}(\b V_{\rm i}-\b V_{\rm e})~.
\end{equation}
and a term proportional to $ (m_{\rm e} nu_{\rm en})/(m_{\rm i}\nu_{\rm in})$ has been neglected.
  The equation of motion of the neutral fluid is:
\begin{eqnarray}
  \rho_{\rm n} \left[ \frac{\partial \b V_{\rm n}}{\partial t}+({\b V_{\rm n}} \cdot\nabla )
{\b V_{\rm n}}\right]\!\!\!&=&\!\!\! -\nabla p_{\rm n}-\nu_{\rm ni}\rho_{\rm n}(\b V_{\rm n}-\b V_{\rm i})-\nonumber\\
&& \nu_{\rm ne}\rho_{\rm n}(\b V_{\rm n}-\b V_{\rm e})+\rho_{\rm n} \b g+\nonumber\\
&&\nabla.\b \tau ~, 
\end{eqnarray}
where $\b g$ is the gravitational acceleration, $\b{\b \tau}$ is the viscous stress tensor of the neutral fluid defined as
\begin{equation}
\tau_{\rm ij}=\mu\left(\frac{\partial V_{\rm ni}}{\partial x_{\rm j}}+ \frac{\partial V_{\rm nj}}{\partial x_{\rm i}}-\frac{2}{3}\delta_{\rm ij}\left(\nabla.\b V_{\rm n}\right)\right),\\i,j=1,2,3. ~.
\end{equation}

 Substituting for $\bm V_{\rm n}-\bm V_{\rm i}$ from Eq. (4), and using $\nu_{\rm in}
\rho_{\rm i}= \nu_{\rm ni} \rho_{\rm n}$ we find:
\begin{eqnarray}
\rho_{\rm n} \left[ \frac{\partial \b V_{\rm n}}{\partial t}+(\b V_{\rm n} 
\cdot\nabla)\b V_{\rm n}\right]=-\nabla p+\frac{{\b J}\times {\b B}}{c}+\rho_{\rm n} \b g+\nonumber\\
 \nabla.\b \tau,~. 
\end{eqnarray}
where $p=p_{\rm n}+p_{\rm i}+p_{\rm e}$ and $\mu$ is the dynamic viscosity.  
Observe that the neutral fluid is subjected to the Lorentz force as a
result of the strong ion-neutral coupling due to their collisions. A comparison with the corresponding equation of V\"{o}gler and Sch\"{u}ssler ( 2007) shows that here the neutral mass density $\rho_n$ appears instead of the total mass density. This is valid for $\rho_n >> \rho_i$ in a weakly ionized plasma. 

Consider Faraday's law of induction:
\begin{equation}
\frac{\partial \b B}{\partial t}=-c\nabla\times\b E
\end{equation}
By substituting for the electric field from Eq.~(2), we get
\begin{equation}
\frac{\partial \b B}{\partial t}=\nabla\times\left[\b V_{\rm e}\times\b B+ \frac{c}{en_{\rm e}}\nabla p_{\rm e}+\frac{m_{\rm e}c\nu_{\rm en}}{e}\left(\b V_{\rm e}-\b V_{\rm n}\right)\right]
~.
\end{equation}

Here $\eta=m_{\rm e}\nu_{\rm en}c^2/(4\pi e^2n_{\rm e})$ is the electrical
resistivity predominantly due to electron-neutral collisions.
Using the construction
\begin{equation}
\b V_{\rm e}\times \b B=[\b V_{\rm n}-(\b V_{\rm n}-\b V_{\rm i})-
                        (\b V_{\rm i}-\b V_{\rm e})]\times\b B~,
\end{equation}
substituting for the relative velocity of the ion and the neutral fluids
from Eq.~(4), assuming $p_{\rm e}=p_{\rm i}$ and neglecting terms of the order $ (m_{\rm e} \nu_{\rm en})/(m_{\rm i}\nu_{\rm in})$ Eq.~(10) becomes:
\begin{eqnarray}
{\partial \b B\over\partial t}=\nabla\times\left[\left(\b V_{\rm n} 
-\frac{\b J}{en_{\rm e}}+\frac{\b J\times\b B}{ c\nu_{\rm in}\rho_{\rm i}}
\right)\times \b{B}\right]+\nonumber\\
\frac{c}{en_{\rm e}}\left(\nabla p_{\rm i}-2\frac{\nabla p_{\rm i}\times\b\Omega_{\rm ic}}{\nu_{\rm in}}\right)+ \nonumber\\
\eta{\nabla}^{2}\b B~.
\end{eqnarray}
where $\b\Omega_{\rm ic}=\frac{e\b B}{m_{\rm i}c}$ is the ion cyclotron frequency.
One can easily identify the Hall term ($\b J/e n_{\rm e}$), and the
ambipolar diffusion term ($\b J\times \b B $) (Brandenburg and Zweibel 1994, Chitre \& Krishan
2001). The Hall term is much larger than the ambipolar term for large
neutral particle densities or for $\nu_{\rm in} \gg \omega_{\rm ci}$.
 In this system
the magnetic field is not frozen to any of the fluids. The induction equation carries the effects of stratification as well as the $\nabla p_{\rm i}\times\b B$ fluid drift. In what follows the electron and the ion densities (pressures) are assumed to be equal and constant. We write the induction equation in a compact form as:
\begin{eqnarray}
{\partial \b B\over\partial t}=\nabla\times\left[\b V_E 
\times \b{B}-\eta\nabla\times\b B\right]~,
\end{eqnarray}
where
\begin{equation}
\b V_{\rm E}= \b V_{\rm n}+\b V_{\rm H}+\b V_{\rm Am}
\end{equation}
with
\begin{eqnarray}
\b V_{\rm H}= -\frac{\b J}{en_{\rm e}}
\end{eqnarray}
as the Hall velocity and 
\begin{eqnarray}
\b V_{\rm Am}= \frac{\b J\times\b B}{ c\nu_{\rm in}\rho_{\rm i}}
\end{eqnarray}
could be called the ambipolar velocity.
One can notice that the Hall effect and the ambipolar diffusion terms are not included in  the induction equation given by V\"{o}gler et al. (2005) and used by V\"{o}gler and Sch\'ussler (2007) to study solar surface dynamo. In order to appreciate the time and the spatial scales over which the additional terms would contribute to the magnetic induction evolution, the induction equation is written in a dimensionless form by using the normalizing quantities $B_0, L_0, t_0, V_0, n_{n0}$ such that $ V_0=L_0/t_0=(B^2_0/4\pi\rho_n)^{1/2}$. The dimensionless induction equation reads:
 
\begin{eqnarray}
\frac{\partial \b B}{\partial t}=\nabla\times\left[\left(\b V_n-\epsilon_H\nabla\times\b B
+\epsilon_{A}\left(\nabla\times\b B\right)\times\b B
\right)\times \b{B}-
 \eta\nabla\times\b B\right],
\end{eqnarray}
where
\begin{eqnarray}
\epsilon_H=\frac{\lambda_i}{L_0}\left(\frac{\rho_n}{\rho_i}\right)^{1/2}~,
 \end{eqnarray}
and $\lambda_i=\frac{c}{\omega_{ip}}$ is the ion inertial scale and $\omega_{ip}=(\frac{4\pi n_i e^2}{m_i})^{1/2}$ is the ion plasma frequency. Thus the Hall term brings in a finite spatial scale in an otherwise ideal MHD system of any arbitrary scale $L$.
The coefficient $\epsilon_A$ of the ambipolar term is found to be: 
\begin{eqnarray}
\epsilon_{A}=\frac{\omega_{ic}}{\nu_{in}}\epsilon_H~.
\end{eqnarray}
clearly highlighting the predominance of the ambipolar diffusion over the Hall effect for large $\b B$ and small $\nu_{in}$.

The mass conservation is described as:
\begin{eqnarray}
	\frac{\partial \rho_{\rm n}}{\partial t}+\nabla\cdot\left(\rho_{\rm n}\b V_{\rm n}\right)=0,\nonumber\\
\nabla\cdot\b V_{\rm i}=0,\nonumber\\
\nabla\cdot\b V_{\rm e}=0~.
\end{eqnarray}

The energy equation for a partially ionized plasma is found to be:
\begin{eqnarray}
\frac{\partial \epsilon}{\partial t}+\nabla\cdot\left[\b V_{\rm n}\left(\epsilon+p+
\frac{\b B^2}{8\pi}\right)\right. \nonumber\\
\left. +\left(V_{\rm H}+ V_{\rm Am}\right)\frac{B^2}{8\pi}-\frac{\left(\b B.\b V_E\right)\b B}{4\pi}\right]= \nonumber\\
-\nabla.\left[\frac{\eta\left(\nabla\times\b B\right)\times\b B}{4\pi}\right]+ \nonumber\\
\left[\nabla.\left(\b V_n.\tau\right)-s_{ij}\tau_{ij}\right]+\nabla.\left(\kappa\nabla T\right)+ \nonumber\\
\rho_n\left(\b g.\b V_n\right)-\frac{4\pi\eta}{c^2}\b J^2+ Q ~.
\end{eqnarray}
where $s_{ij}=\frac{\partial V_{\rm ni}}{\partial x_{\rm j}}$. Here $Q$ includes the rest of the nonadiabatic effects such as the radiative cooling. The energy equation explicitly displays the additional terms due to the Hall and the ambipolar effects. The effect of the ionization potential $\chi_i$ has not been taken explicitly into account in the internal energy. Instead one could continue with the methodology used by V\"{o}gler et al. (2005) to consider pressure $p$ and temperature $T$ to be a function of the internal energy .

Equations (8), (13), (20) and (21) describe the solar surface dynamo.

\section{The alpha effect in compressible three-component turbulent magnetofluid}
The $\alpha$ effect in an incompressible partially ionized plasma has been discussed recently ( Krishan and Gangadhara 2008). Here an attempt is made to include the effect of turbulent density fluctuation of the neutral fluid along with the usual velocity and magnetic field fluctuations. The ambipolar diffusion term depends on the density of the neutral fluid through the ion-neutral collision frequency. The electron and the ion densities are kept constant and their fluctuations are ignored. Following the standard
procedure (Krause \& R\"adler 1980) the velocity $\b V_{\rm E}$, the
magnetic field $\b B$ and the neutral fluid density $\rho_n$ are split into their average large scale parts
and the fluctuating small scale parts as:
\begin{eqnarray}
\b V_{\rm E} &=& \oline{\b V_{\rm E}}+\b V'_{\rm E}, \\  
       \b B &=& \oline{\b B}+\b B', \\ \rho_n &=& \oline{\rho_n }+\rho_n'
\end{eqnarray}
such that
\begin{eqnarray}
\oline{\b V'_{\rm E}}=0, \quad\quad \oline{\b B'}=0, \quad\quad \oline{\rho_n'}=0.
\end{eqnarray} 
In the mean-field dynamo the magnetic induction equation is solved for
large and small scale fields. Here, the mass conservation equation of the neutral fluid would also be considered.  Substituting Eqs.~(22-25) into
the induction equation (13), we find, in the first order smoothing
approximation,
\begin{eqnarray}
\b V'_{\rm E}=\b V'_n-\frac{\b J'}{en_{\rm e}}+\frac{\b J'\times\oline {\b B}}
{c\nu_{\rm in}\rho_{\rm i}}
+ \frac{\oline{\b J}\times\b B'}{c\nu_{\rm in}\rho_{\rm i}}-\frac{\nu_{in}'}{c\rho_i \oline{\nu_{in}}^2}\oline{\b J}\times \oline{\b B}~.
\end{eqnarray}
and the mean flow is found to be:
\begin{eqnarray}
\oline{\b V_{\rm E}}=\oline{\b V_{\rm n}}-\frac{\oline {\b J}}{en_{\rm e}}+
             \frac{\oline{\b J}\times\oline{\b B}}{c\oline{\nu_{\rm in}}\rho_{\rm i}}+
             \frac{\oline{\b J'\times\b B'}}{c\oline{\nu_{\rm in}}\rho_{\rm i}}-\frac{\oline{\oline{\b J}\times \b B'\nu_{in}'}}{c\rho_i \oline{\nu_{in}}^2} \nonumber\\
-\frac{\oline{\b J'\times\oline{ \b B}\nu_{in}'}}{c\rho_i \oline{\nu_{in}}^2}~.
\end{eqnarray}
The fluctuation in the ion-neutral collision frequency is defined as:
\begin{eqnarray}
\nu_{in}'=A\rho_n',~.
\end{eqnarray}
\begin{eqnarray}
 A=\frac{\Sigma_{in}}{m_n}\left(\frac{8k_BT}{\pi m_{in}}\right)^{1/2}~.
\end{eqnarray}
where the cross-section $\Sigma_{in}\sim 5\times 10^{-15}$ cm$^2$ and $m_{in}=\frac{m_im_n}{m_i+m_n}$ is the effective mass of the scattering particles.

The turbulent electromotive force $\Xi$ is a function of the mean
magnetic induction $\oline{\b B}$ and mean quantities formed from the
fluctuations, and is expressed as:
\begin{eqnarray}
\Xi =\oline{\b V'_{\rm E}\times\b B'} = \alpha \oline{\b B} - 
     \beta\, \nabla\times\oline{\b B}~,
\end{eqnarray}
where
\begin{eqnarray}
\alpha &=& - \frac{\tau_{\rm cor}}{3}\oline{\b V'_{\rm E}\cdot
                  (\nabla\times\b V'_{\rm E})}\nonumber\\
       &=& \alpha_{\rm v}+\alpha_{\rm H}+\alpha_{\rm Am}~.
\end{eqnarray}
Here
\begin{eqnarray}
\alpha_{\rm v} = -\frac{\tau_{\rm cor}}{3}\oline{\b V'_{\rm n} \cdot\Omega'_{\rm n}}
\end{eqnarray}
is the measure of the average kinetic helicity of the neutral fluid in
the turbulence possessing correlations over time $\tau_{cor}$. Retaining only the first order contributions from the Hall and the ambipolar effects, the contribution to $\alpha$ from the Hall effect is found to be 
\begin{eqnarray}
\alpha_{\rm H} = \frac{2\tau_{\rm cor}}{3en_{\rm e}}\oline{\b J'
                 \cdot\b\Omega'_{\rm n}}
\end{eqnarray}
 The coupling of the
charged components with the neutral fluid is clearly manifest through
the possible correlation between the current density fluctuations and
the vorticity fluctuations of the neutral fluid 
$\b \Omega'_{\rm n}=\nabla\times \b V'_{\rm n} .$ The ambipolar term gives 
rise to  
\begin{equation}
\alpha_{\rm Am} = \b \alpha_{\rm A_0}\cdot\oline{\b B}+\b \alpha_{\rm A_1}\cdot\oline{\b J}+\b \alpha_{\rm A_2}\cdot\oline{\b J}\times\oline{\b B}~,
\end{equation}
with
\begin{equation}
\b \alpha_{\rm A_0} = \frac{2\tau_{\rm cor}}{3c\rho_{\rm i}\oline{\nu_{\rm in}}}\oline
                    {\b J'\times\b\Omega'_{\rm n}}~,
\end{equation}
\begin{equation}
\b \alpha_{\rm A_1} = -\frac{2\tau_{\rm cor}}{3c\rho_{\rm i}\oline{\nu_{\rm in}}}\oline
                    {\b B'\times\b\Omega'_{\rm n}}~,
\end{equation}
and
\begin{equation}
\b \alpha_{\rm A_2} =\left(\frac{-A}{\oline{\nu_{in}}}\right) \frac{2\tau_{\rm cor}}{3c\rho_{\rm i}\oline{\nu_{\rm in}}}\oline{\b V_n'\times\nabla \rho_n'}~.
\end{equation}
The contributions to $\alpha$ from the ambipolar diffusion with its essential
nonlinear character manifest through its dependence on the average
magnetic induction and its spatial variation. Note that the term $\b \alpha_{\rm A_1}$ was ignored in Krishan and Gangadhara (2008) as therein only $\oline{\b B}$ dependent terms were retained. Now that the terms proportional to $\oline{\b J}\times\oline{\b B}$ appear, it becomes necessary to retain terms proportional to $\oline{\b J}$.
The compressibilty effect appears through the correlation between the gradient of the density fluctuation and the velocity fluctuation and this term brings in a quadratic dependence on the large scale field $\oline{\b B}$. One also observes that the Hall alpha (Eq. 33)
requires a component of the fluctuating current density along the
fluctuating vorticity of the neutral fluid whereas the ambipolar
effect (Eq. 35) thrives on the component of the fluctuating current
density perpendicular to the fluctuating vorticity. 
The turbulent
dissipation parameter $\beta$ is given by
\begin{eqnarray}
\beta = \frac{\tau_{\rm cor}}{3}\oline{{\b V}'^2_{\rm E}}
= \beta_{\rm v}+\beta_{\rm H}+\beta_{\rm Am}
\end{eqnarray}
with
\begin{eqnarray}
\beta_{\rm v} = \frac{\tau_{\rm cor}}{3}\oline{{\b V}'^2_{\rm n}}
\end{eqnarray}
as the measure of the average turbulent kinetic energy of the neutral
fluid in the turbulence possessing correlations over time
$\tau_{\rm cor}$. The first order Hall contribution to $\beta$ is
\begin{eqnarray}
\beta_{\rm H} =- \frac{2\tau_{\rm cor}}{3en_{\rm e}}\oline{\b J' \cdot\b V'_{\rm n}}
\end{eqnarray}
 The coupling of the
charged components with the neutral fluid is clearly manifest through
the possible correlation between the current density fluctuations and
the velocity fluctuations of the neutral fluid. The ambipolar term furnishes
\begin{equation}
\beta_{\rm Am} = \b \beta_{A0}\cdot \oline{\b B}+ \b \beta_{A1}\cdot \oline{\b J}+ \b \beta_{A2}\cdot\left( \oline{\b J}\times\oline{\b B}\right) ~,
\end{equation}
where
\begin{equation}
\b \beta_{\rm A0} = \frac{2\tau_{\rm cor}}{3c\rho_i\oline{\nu_{\rm in}}}\oline{\b 
                    V_n'\times\b J'}~,
\end{equation}
\begin{equation}
\b \beta_{\rm A1} = \frac{2\tau_{\rm cor}}{3c\rho_i\oline{\nu_{\rm in}}}\oline{\b 
                    V_n'\times\b B'}~,
\end{equation}
\begin{equation}
\b \beta_{\rm A2} = \left(\frac{-A}{\oline{\nu_{in}}}\right)\frac{2\tau_{\rm cor}}{3c\rho_i\oline{\nu_{\rm in}}}\oline{\b 
                    V_n' \rho_n'}~.
\end{equation}
with its essential
nonlinear character manifest through its dependence on the average
magnetic induction $\oline{\b B}$, the average current density $\oline{\b J}$ and the average Lorentz force $\oline{\b J}\times \oline{\b B}$ . One also observes that the Hall $\beta_{\rm H}$ requires
a component of the current density fluctuations along the velocity
fluctuations of the neutral fluid whereas the ambipolar effect thrives
on the component of the current density fluctuations and the magnetic field fluctuations perpendicular to
the velocity fluctuations in addition to the velocity- density correlation. We have used rigid or perfectly conducting
boundary conditions (all surface contributions vanish) while
determining the averages. Here we consider what is known as the
$\alpha^2$ dynamo and take the mean flow $\oline{\b V_{\rm E}}=0$. This
actually determines the relative mean flow amongst the three
fluids. The dynamo equation reduces to
\begin{equation}
{\partial \b B\over\partial t}=\nabla\times\left[\alpha{\b B}
-\beta \nabla\times {\b B}\right]+ \eta{\nabla}^{2}{\b B}~.
\end{equation}
 The large scale magnetic field is written without the bar. The solutions of the dynamo equation for some representative cases were discussed in Krishan and Gangadhara ( 2008). Here the dynamo equation will be set up for the case where the neutral density fluctuations, through the ambipolar diffusion,  are the major contributors to the $\alpha$ effect. Thus the dynamo equation becomes:
\begin{equation}
{\partial \b B\over\partial t}=\nabla\times\left[\left(\b {\alpha_{\rm A_2}}\cdot \b J\times \b B\right){\b B}
-\beta_{\rm A2} \nabla\times {\b B}\right]+ \eta{\nabla}^{2}{\b B}~.
\end{equation}
The equation  even in this reduced form is still quite complicated. In order to have a glimpse of the role of the ambipolar term through the density fluctuations, we ignore the dissipation term  and estimate the characteristic timescale of evolution of the field to be: 
\begin{equation}
t\approx \frac{L}{\left(\b {\alpha_{\rm A_2}}\cdot \b J\times \b B\right)}~.
\end{equation}
where
\begin{equation}
 \alpha_{\rm A_2} =\frac{-2\tau_{\rm cor}\oline{\b V_n'\times\nabla \rho_n'}}{4\pi \oline{\rho_n}\rho_i \oline{\nu_{in}}}~.
\end{equation}

The velocity-density correlation may be approximated  by using the steady state mass conservation
\begin{equation}
\nabla\cdot\left[\left(\oline{\rho_n}+\rho_n'\right)\left(\oline{ \b V_n}+ \b V_n'\right)\right]=0~.
\end{equation}
Taking the average and assuming $\|\oline{\b V_n'\cdot\nabla\rho_n'}\|\approx \|\oline{\rho_n'\nabla\cdot\b V_n'}\|$, $\|\oline{\b V_n}\cdot\nabla\oline{\rho_n}\|\approx \|\oline{\rho_n}\nabla\cdot\oline{\b V_n}\|$    and $\|\oline{\b V_n'\cdot\nabla\rho_n'}\approx \oline{\b V_n'\times\nabla\rho_n'}\|$;
 
for the case of $\oline{\b V_E}=0$, retaining only the non-turbulent contribution,
$\oline{\b V_n} = \frac{\oline{\b J}}{en_e}-\frac{\oline{\b J}\times\oline{\b B}}{c\oline{\nu_{in}}\rho_i}$.
the characteristic timescale is found to be:
\begin{equation}
t\approx \tau_{cor}\frac{t_A^4\nu_{in}^2}{\tau_{cor}^2}\frac{\rho_i^2}{\rho_n^2}~.
\end{equation}
where $t_A=\frac{L}{V_{nA}}$ is the Alfven crossing time of the characteristic spatial scale $L$ and $V_{nA}=\frac{B}{\surd{4\pi\oline{\rho_n}}}$ is the Alfven speed of the neutral fluid. For the collisional plasma $t_A \nu_{in} > 1$ and $t_A$ is also expected to be greater than $ \tau_{cor}$ as well. Thus depending on the ion to neutral density ratio the magnetic field evolution may take place on time scale as short as $\tau_{cor}$. This rather crude estimate of the time scale is at best a pointer towards a possible rapid evolution of the magnetic field due to ambipolar diffusion on the solar atmosphere.

\section{Conclusion}
 
In this first attempt at formulating the mean-field dynamo in a compressible partially ionized plasma, a complete set of multi-fluid- magnetic equations, in particular the Faraday law and the energy equations have been established with the hope that adequate computational resources would be deployed to understand this very important problem of the solar surface dynamo. Additional physics of the Hall and the ambipolar diffusion effects brings in a variety of spatial and time scales associated with the ion inertial scale, the gyrofrequencies and the collisional frequencies, the scales towards which the high resolution observations of the solar magnetic fields tend to lead us.

\section*{Acknowledgments}
The author acknowledges the support received from Dr. Varghese for the preparation of this manuscript. 

\label{lastpage}
\end{document}